\documentclass[prl,twocolumn,a4paper,showpacs,superscriptaddress]{revtex4-1}
\usepackage[]{fontenc}
\usepackage[latin9]{inputenc}
\usepackage{color}
\usepackage{amsmath}
\usepackage{graphicx}
\usepackage{amssymb}
\usepackage{epstopdf}

\makeatletter

\providecommand{\tabularnewline}{\\}

\@ifundefined{textcolor}{}
{%
 \definecolor{BLACK}{gray}{0}
 \definecolor{WHITE}{gray}{1}
 \definecolor{RED}{rgb}{1,0,0}
 \definecolor{GREEN}{rgb}{0,1,0}
 \definecolor{BLUE}{rgb}{0,0,1}
 \definecolor{CYAN}{cmyk}{1,0,0,0}
 \definecolor{MAGENTA}{cmyk}{0,1,0,0}
 \definecolor{YELLOW}{cmyk}{0,0,1,0}
 }

\makeatother

\begin{document}
\title{\noindent \textit{Ab initio} investigation of Elliott-Yafet electron-phonon
mechanism in laser-induced ultrafast demagnetization}

\author{K. Carva}
\email{karel.carva@fysik.uu.se}
\affiliation{Department of Physics and Astronomy, Uppsala University,
Box 516, S-75120 Uppsala, Sweden}
\affiliation{Charles University, Faculty of Mathematics and Physics,
  Department of Condensed Matter Physics,
  Ke Karlovu 5, CZ-12116 Prague 2, Czech Republic}
\author{M. Battiato}
\author{P.\,M. Oppeneer}
\affiliation{Department of Physics and Astronomy, Uppsala University,
Box 516, S-75120 Uppsala, Sweden}
\date{\today}
\begin{abstract}
\noindent 
The spin-flip (SF) Eliashberg function is calculated from first-principles for ferromagnetic Ni to accurately establish the 
contribution of Elliott-Yafet electron-phonon SF scattering to Ni's femtosecond laser-driven demagnetization. 
This is used to compute the SF probability and demagnetization rate for laser-created thermalized as well as non-equilibrium electron distributions.
Increased SF probabilities are found for thermalized electrons, but the induced demagnetization rate is extremely small.  A larger 
demagnetization rate is obtained for {non-equilibrium} electron distributions, but its contribution is too small to account for
femtosecond 
demagnetization.
\end{abstract}
\maketitle

Ultrafast demagnetization of ferromagnetic metals through excitation by a femtosecond laser pulse was discovered  fifteen years ago by Beaurepaire {\it et al.} \cite{beaurepaire96}.
In spite of intensive investigations the microscopic origin of the ultrafast demagnetization could not be disclosed and continues to be controversially debated (see \cite{kirilyuk10} for a recent review). Several mechanisms have been proposed to explain the observed 
ultrafast phenomenon 
 \cite{zhang00,koopmans05,carpene08,krauss09,bigot09,koopmans10,atxitia10,battiato10,schmidt10}.
Most of these theories assume the existence of an ultrafast spin-flip (SF) channel, which would cause dissipation of spin angular momentum within a few hundred femtoseconds.  

Elliott-Yafet electron-phonon SF scattering has been proposed as a mechanism for  ultrafast spin-dissipation \cite{koopmans05}. Strong support in favor of electron-phonon mediated spin-flips as the actual mediator of the femtosecond demagnetization was made in a very recent work, in which {\it ab initio} calculated SF probabilities for thermalized electrons compared favorably to  SF probabilities derived from pump-probe demagnetization measurements \cite{koopmans10}. While these results definitely favor the Elliott-Yafet SF scattering mechanism, the calculation of the electron-phonon scattering involved several serious approximations.  Applying the so-called Elliott approximation \cite{elliott54} only spin-mixing due to spin-orbit coupling in the {\it ab initio} wavefunctions was included, but no electron-phonon matrix elements and no real phonon dispersion spectrum was considered. The thus-obtained SF probability is however not a direct measure of demagnetization. 
Recent model simulations for thermalized hot electrons \cite{atxitia10} using the Landau-Lifshitz-Bloch equation \cite{atxitia07} and assuming a fitted SF parameter did reproduce the experimental magnetization response, but couldn't assign the SF origin.
Hence, it remains a crucial, open question whether laser-induced demagnetization can 
indeed be attributed to electron-phonon mediated SF scattering.

Here we report an {\it ab initio} investigation to accurately establish the extent to which the Elliott-Yafet electron-phonon SF scattering contributes to fs demagnetization. To this end we perform {\it ab initio} calculations for ferromagnetic Ni, which ultrafast magnetization decay is well documented \cite{beaurepaire96,stamm07,koopmans10}.
We include the full electron-phonon matrix elements and phonon dispersions in our calculations. 
Introducing an energy-dependent SF Eliashberg function we compute SF probabilities and demagnetization rates for laser-heated thermalized electrons as well as laser-induced non-equilibrium electron distributions, from which we draw qualified conclusions on the possibility of phonon-mediated demagnetization.  

To treat phonon-mediated 
SF scattering at variable electron energies we define a \textit{generalized} energy- and spin- dependent Eliashberg function,
\begin{eqnarray}
 \alpha_{\sigma\sigma'}^{2}F(E,\Omega)  = \frac{1}{2M\Omega}\sum_{\nu,n,n'}\iint d\mathbf{k} d\mathbf{k}' 
g_{\mathbf{k}n,\mathbf{k}'n'}^{\nu \sigma\sigma'} (\mathbf{q}) \times \nonumber \\
 \, \,\, \,  \delta (\omega_{\mathbf{q\nu}}- | \Omega | )\delta (E_{\mathbf{k}n}^{\sigma}-E )\delta (E_{\mathbf{k'}n'}^{\sigma'}-E)\,,
\label{eq:SFElbgEn} 
\end{eqnarray}
which comprises initial and final electron states with quantum numbers $\mathbf{k}n$, $\mathbf{k}'n'$ that interact through a phonon with frequency $\Omega$=$\omega_{\mathbf{q\nu}}$,  $\nu$ and $\bf q$ denote its mode and wavevector.
$M$ is the ionic mass,  $\sigma$=$\uparrow,\downarrow$ denote the spin majority, miniority components.
For $E$=$E_F$ (the Fermi energy) the SF part $\alpha_{\uparrow\downarrow}^{2}F(E_F,\Omega)$ gives the SF Eliashberg function \cite{fabian99} and the sum over all $\sigma\sigma'$ corresponds to the standard Eliashberg function, $\alpha^{2}F(E_F,\Omega)$ \cite{grimvall81}.
The (squared) electron-phonon matrix elements are 
\begin{equation}
g_{\mathbf{k}n,\mathbf{k}'n'}^{\nu \sigma\sigma'}(\mathbf{q} )=| \mathbf{u}_{\mathbf{q}\nu}\cdot  \langle\Psi_{\mathbf{k}n}^{\sigma} |\nabla_{\mathbf{R}}V | \Psi_{\mathbf{k}'n'}^{\sigma'} \rangle |^{2}\,,
\label{eq:ElPhonElem}
\end{equation}
where  $V$ is the 
potential, $\mathbf{u}_{\mathbf{q}\nu}$ the phonon polarization vector and  $| \Psi_{\mathbf{k}n}^{\sigma}\rangle$
are the eigenstates in the ferromagnet.
Momentum conservation requires $\mathbf{q=k'-k}$.
SF scattering becomes possible through the relativistic spin-orbit coupling. The majority, minority Bloch states 
 $| \Psi_{\mathbf{k}n}^{\uparrow}\rangle$ and $| \Psi_{\mathbf{k}n}^{\downarrow}\rangle$ can be decomposed in pure spinor components
\begin{equation}
| \Psi_{\mathbf{k}n}^{\uparrow}\rangle = a_{\mathbf{k}n}^{\uparrow} ( {{}_1 \atop {}^0} ) +b_{\mathbf{k}n}^{\uparrow} ( {{}_0 \atop {}^1} ),\; |\Psi_{\mathbf{k}n}^{\downarrow}\rangle = a_{\mathbf{k}n}^{\downarrow} ( {{}_0 \atop {}^1} ) +b_{\mathbf{k}n}^{\downarrow}( {{}_1 \atop {}^0} ) ,
\label{eq:EigDecomp}
\end{equation}
where the components $b_{\mathbf{k}n}^{\sigma}$ are nonzero only
if spin-orbit coupling is present and represent the degree of spin-mixing, which is a precondition for nonzero $g_{\mathbf{k}n,\mathbf{k}'n'}^{\nu\uparrow \downarrow}$.

To study demagnetization we consider two quantities, SF probabilities and spin-resolved transition rates. The latter are defined as \cite{yafet63}
\begin{eqnarray}
S^{\sigma\sigma'}\!\!=\!\!\iint\!\!\alpha_{\sigma\sigma'}^{2}F(E,\Omega)f_{\sigma}(E)(1\!-\!f_{\sigma'}(E\!+\!\hbar \Omega)) \times \,\,\,\,  \nonumber \\
\, \,\, \,\,\,
(\Theta(\Omega)\!+\!N(\Omega))\ d\Omega dE.
\label{eq:SpinEvGen0}
\end{eqnarray} 
Here $N (\Omega)$ is the phononic Bose-Einstein distribution, $f_{\sigma}$ the Fermi distribution, and $\Theta( \Omega) $ the Heaviside function.
Important for the effective demagnetization is the spin decreasing rate $S^{-}$, which corresponds to $S^{\uparrow\downarrow}$, while the increasing one  $S^{+}$ corresponds 
to $S^{\downarrow\uparrow}$.

An approximation of Eq.\ (\ref{eq:SpinEvGen0}) is helpful to achieve a faster evaluation and provide more insight in the process.
Energy conservation during electron-phonon scattering requires $E_{\mathbf{k'}n'}-E_{\mathbf{k}n}\!=\!\hbar\Omega $, but the phonon energy $\hbar\Omega$ is usually very small ($<\! 0.04$ eV) compared to electron related properties. Already in the standard Eliashberg formulation Eq.\ 
(\ref{eq:SFElbgEn}) an energy difference between initial and final states is neglected while the $\delta$-functions $\delta (E_{\mathbf{k}n}^{\sigma}-E )$ are broadened with a parameter (0.03 eV, here). 
Similarly, one can neglect the energy variation due to $\hbar\Omega$ in the Fermi function
$f_{\sigma}(E+\hbar \Omega)$, as long as the temperature is high enough.
We can then rewrite spin-resolved transition rates in the form 
\begin{equation}
S^{\sigma\sigma'}=\int w_{\sigma\sigma'}(E)f_{\sigma}(E)(1-f_{\sigma'}(E))dE,
\label{eq:SpinEvGen}
\end{equation}
where we introduced the energy- and spin- dependent specific scattering rate for electrons 
$w_{\sigma\sigma'}$ given by
\begin{equation}
w_{\sigma\sigma'}\!\left(E\right)\!= \int_{0}^{\infty}d\Omega \,\alpha_{\sigma\sigma'}^{2}F (E,\Omega) (1\!+2\! N\! (\Omega ) ) .
\label{eq:SFRate_En}
\end{equation}
Note that $w_{\uparrow\downarrow}(E)\! =\! w_{\downarrow\uparrow}(E)$.
All calculations were checked against a more accurate numeric implementation not involving this approximation.
The SF probability for an electron with energy $E$ is defined as the ratio of the SF part to the corresponding total counterpart, $p_{S}(E)=$ $2w_{\uparrow\downarrow} (E)/\sum_{\sigma\sigma'}w_{\sigma\sigma'}(E)$. 
Analogously, the total SF probability during a scattering event can be defined as
\begin{equation}
P_S=(S^{-}+S^{+})/\sum_{\sigma\sigma'}S^{\sigma\sigma'} . 
\label{eq:TotalProb}
\end{equation}
Although the SF probability has been used in recent discussions of laser induced-demagnetization \cite{steiauf09,koopmans10},
it is actually not the crucial quantity (as a high but equal SF probability for both spin channels would not cause a demagnetization). We define therefore the normalized demagnetization ratio,
$D_S=(S^{-}-S^{+})/\sum_{\sigma\sigma'}S^{\sigma\sigma'}$, 
which tracks the difference of magnetic moment increasing and decreasing SF contributions. 


To investigate phonon-induced demagnetization 
in laser-excited Ni we proceed now in three steps.
First, we compute the {\it ab initio} SF probability $P_S$ for equilibrium Ni, i.e., for $E\!=\!E_F$. 
Second, we compute SF probabilities $P_{S}$ for laser-heated Ni, by treating a range of electron energies that correspond to those in a hot, thermalized electron gas after laser-excitation. Thermalization to electron temperatures $T_e$ of a few thousand K occurs quickly within about 200 fs after the laser pulse, but the hot electrons are not in equilibrium with the lattice and the lattice temperature is not altered significantly. 
In the third step we consider the SF probability for non-equlibrium (NEQ) electron distributions
\cite{oppeneer04} that are expected to be present within $\sim$100 fs after laser stimulation. Demagnetization ratios $D_S$ are subsequently evaluated for these three situations. 
The results obtained in these steps are furthermore compared to values which we compute with the so-called Elliott relation (see below). 

An {\it ab initio} evaluation of the SF probability of equilibrium Ni
requires calculated phonon dispersions and a relativistic electronic structure. 
Such calculation has previously been done for paramagnetic Al \cite{fabian99}, but has not yet
been accomplished for ferromagnets. An approximation was introduced years ago by 
Elliott \cite{elliott54}, who pointed out a possible source of 
SF scattering arising from the spin-mixing of eigenstates.
Employing several assumptions, {\it viz}.\ a paramagnetic metal, nearly constant electron-phonon matrix elements, $b_{\mathbf{k}n}$ constant in the Brillouin zone, and $b_{\mathbf{k}n}^{\sigma}\ll$ $ a_{\mathbf{k}n}^{\sigma}$, Elliot derived a relation between the spin lifetime $\tau_{S}$ for a general kind of
scattering event with lifetime $\tau$. 
This so-called Elliott relation 
uses the Fermi surface averaged spin-mixing of eigenstates
$\langle b^{2} \rangle =$ $\sum_{\sigma,n}\int d{\mathbf{k}}  | b_{\mathbf{k}n}^{\sigma} |^{2} {\delta} (E_{\mathbf{k}n}^{\sigma}-E_{F} )$
and predicts the SF probability $P_S^{b^2} = (\tau_S/\tau)^{-1} = 4 \langle b^{2} \rangle$.

In a similar way as introduced above, the influence of spin-mixing on the SF probability in laser-heated Ni can be evaluated.
We define a SF density of states (DOS) as 
\begin{equation}
n_{\uparrow\downarrow}(E) = \sum_{n,\sigma}\int d{\mathbf{k}} 
| b_{\mathbf{k}n}^{\sigma} |^{2}
 {\delta} (E_{\mathbf{k}n}^{\sigma}-E).
\label{eq:EllRelDOS}
\end{equation}
A generalized Elliott SF probability for an electron with energy $E$ is then given as $P_{S}^{b^{2}}(E)=4 n_{\uparrow\downarrow} (E)/$ $n (E)$ (with $n (E)$ the total DOS) which yields the standard Elliott expression  $\langle b^{2} \rangle$ in the limit $b_{\mathbf{k}n}^{\sigma}\ll a_{\mathbf{k}n}^{\sigma}$ and  $E\!=\!E_F$. The total SF probability $P_{S}^{b^{2}}$ of a laser-heated system with electron distribution $f_{\sigma}(E)$ is obtained from Eqs.\ (\ref{eq:TotalProb}) and (\ref{eq:SpinEvGen}), where $w_{\uparrow\downarrow}(E)$ is replaced by $n_{\uparrow\downarrow}(E)$ and $w(E)$ by $n(E)$.
Note that although the treatment is intended for phonon scattering the Elliott relation in fact does not take the character of scattering involved into account.
Also, the assumption of a paramagnetic material is essential in Elliott's derivation as this permits SF scattering {\it in each $\bf k$ point} in the spin-degenerate majority, minority bands at $E_F$.  
Experimentally the Elliott relation was found to be valid up to a multiplication
by a material specific constant with variation smaller than one order
of magnitude for various paramagnetic metals \cite{beuneu78}. 
Recently it has also been applied to ferromagnetic metals \cite{steiauf09,koopmans10},  even though for exchange-split
ferromagnetic bands there exist far less $\bf k$ points at which spin-degenerate bandcrossings occur.

%
\begin{figure}[tb]
\includegraphics[clip,width=0.8\columnwidth]{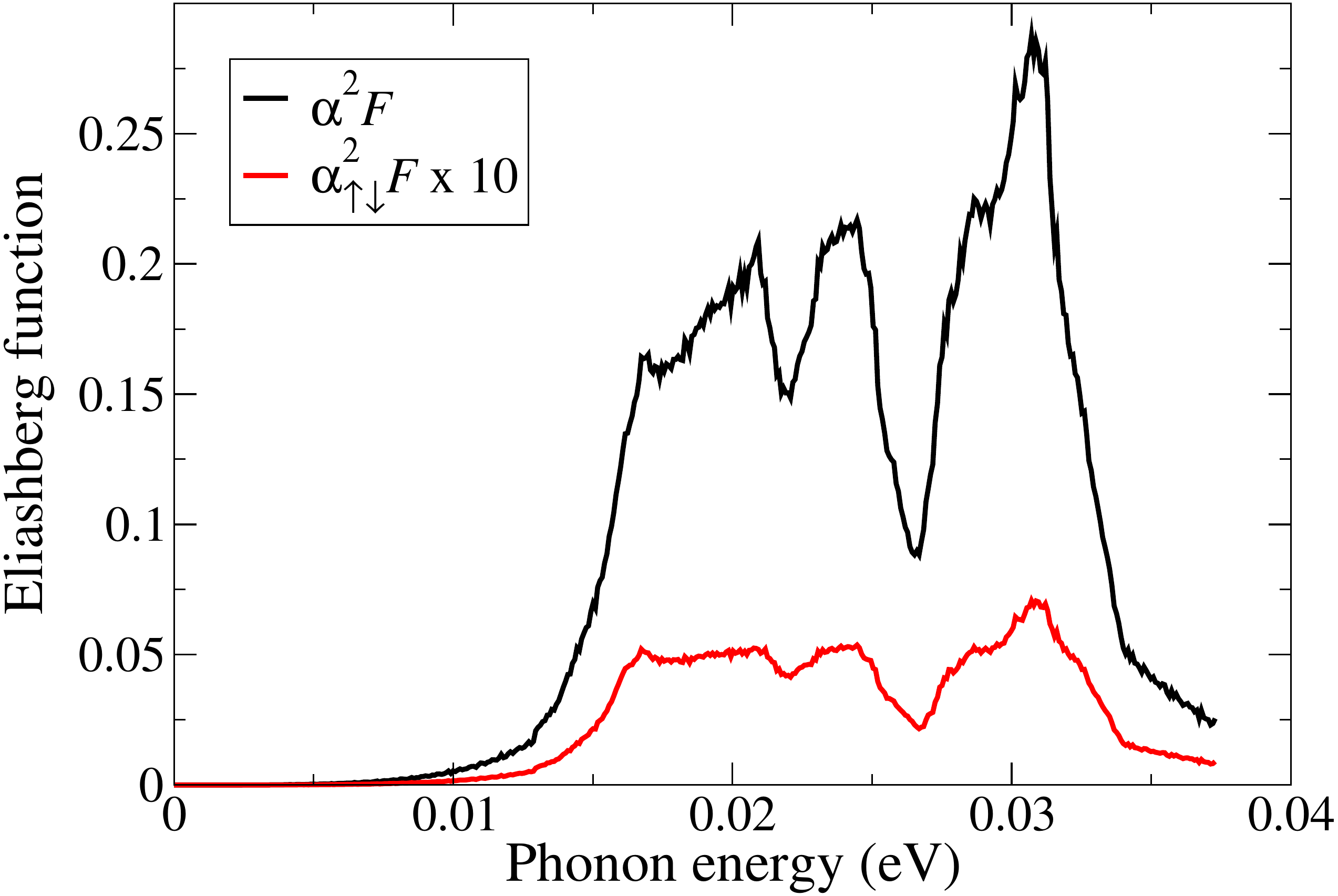}
\caption{(Color online) {\it Ab initio} calculated Eliashberg $\alpha^2 F(E_F,\Omega)$ and SF Eliashberg $\alpha^2_{\uparrow\downarrow} F(E_F,\Omega)$
 functions of Ni in equilibrium.}
\label{fig:Elbg_SFElbg_Ni-1} 
\end{figure}

We have tested the implementation by computing first Al and Ni in equilibrium
at low temperature (\textless{}$300$\,K).
Our calculations are based on the density functional theory (DFT) within
the local spin-density approximation (LSDA), see \cite{details} for details.
For Al our calculated $\alpha^2_{\uparrow\downarrow} F$ is of the order of $10^5$ smaller than $\alpha^2 F$ and in good agreement with the existing previous result \cite{fabian99}.
The {\it ab initio} calculated SF and non-SF Eliashberg functions of equilibrium Ni are shown in Fig.\ \ref{fig:Elbg_SFElbg_Ni-1}.
For Ni the computed  SF $\alpha^2_{\uparrow\downarrow} F$ function is only about 50 times smaller than the ordinary $\alpha^2 F$ function; this is due to the larger spin-orbit coupling.
The resulting total SF probability, $P_S$=0.04, is given in Table \ref{tab:SFRates-1}.
To estimate the accuracy of the Elliott approximation we have
calculated the Elliott SF probability and obtain $P_{S}^{b^{2}}$=$0.07$.
This value is in rough agreement with $P_{S}^{b^{2}}$=$0.10$  computed in Ref.\ \cite{steiauf09}. Thus we find that the Elliott relation overestimates the SF probability in equilibrium Ni by about a factor two.

\begin{table}[b]
\begin{ruledtabular}
\caption{Calculated spin-flip probabilities $P_S$ and demagnetization ratios $D_S$ for laser-pumped Ni. Results are given for equilibrium (low $T$), for thermalized electrons at a high Fermi temperature $T_e$, and for the non-equilibrium (NEQ) electron distribution created by fs laser-excitation.   Results obtained for the approximate Elliott SF probability $P_{S}^{b^{2}}$ (this work and \cite{steiauf09}) are given for comparison.
 }
\begin{tabular}{l c c c c}
  & $P_S^{b^{2}}$ & $P_{S}$ & $D_S$  \tabularnewline
\hline 
Ni (low $T$) & 0.07 
(0.10 \cite{steiauf09}) 
& $0.04$ & 0 \\
Ni ($T_{e}\!=\!1500$K) & 0.08 & 0.05 & 0.002\tabularnewline
Ni ($T_{e}\!=\!3000$K) & 0.11 & 0.07 & 0.003\tabularnewline
Ni ($T_{e}\!=\!5000$K) & 0.12 & 0.10 & 0.004\tabularnewline
Ni (NEQ) & 0.12 & 0.09 & 0.025 \tabularnewline
\end{tabular}
\label{tab:SFRates-1}
\end{ruledtabular}
\end{table}

\begin{figure}[tb]
\includegraphics[clip,width=0.85\columnwidth]{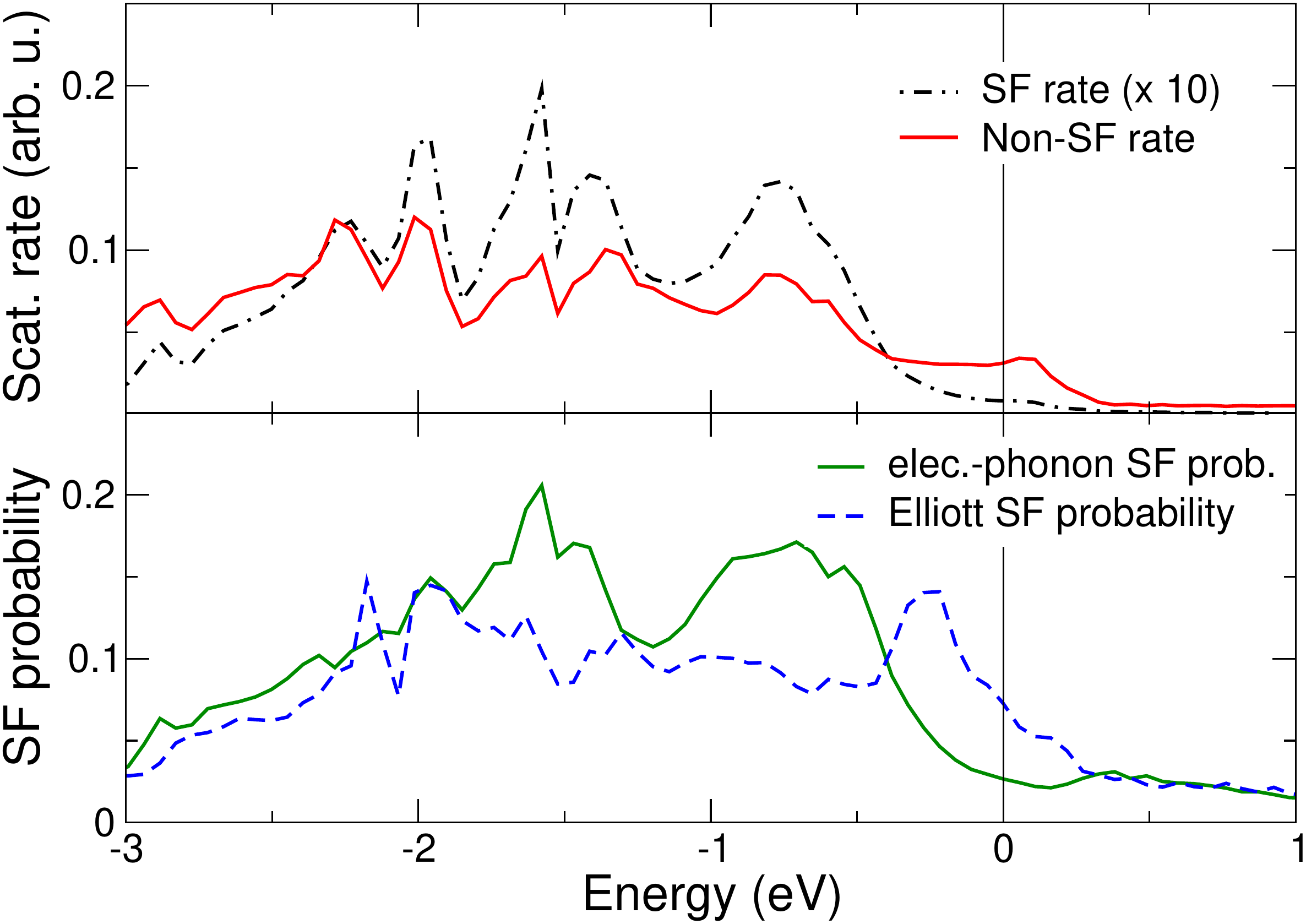}
\caption{(Color online) Energy-resolved electron-phonon total and SF
scattering rates $w (E)$ and $w_{\uparrow\downarrow} (E)$ of Ni, and normalized SF probability
 $P_{S} (E)$ and approximate SF probability $P_{S}^{b^{2}}(E)$ obtained
 from the Elliott relation.}
\label{fig:En_SF}
\end{figure}

\begin{figure}[tb]
\includegraphics[angle=90,bb=35bp 60bp 580bp 790bp,clip,width=0.7\columnwidth]{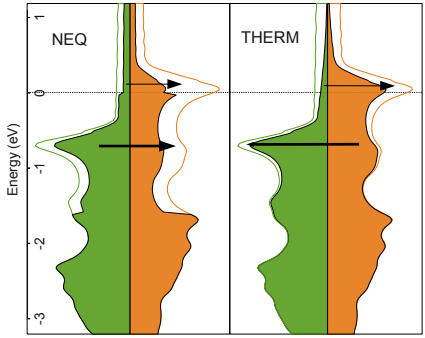}
\caption{(Color online) Spin-resolved 
DOS (filled areas) and phonon induced spin-flips (arrows) of NEQ and electron thermalized Ni. The equilibrium DOS is shown
by thin lines. SF transitions are significantly
different at energies above and below $E_{F}$ (=0 eV). 
The arrows thickness corresponds to the
transition rate, its direction and length give
which direction is dominant and how much. 
The amount of laser redistributed electrons has been enlarged to improve visibility. }
\label{fig:TransScheme}
\end{figure}

Next we turn to the topic of current controversy, the actual amount of phonon-induced
demagnetization in laser-excited Ni. 
In Fig.\ \ref{fig:En_SF}(top) we show calculated energy-resolved SF and non-SF scattering rates ($w_{\uparrow\downarrow}(E)$ and  $w (E)$).
Note the strong energy variations of $w (E)$.
In Fig.\ \ref{fig:En_SF}(bottom) we compare the computed electron-phonon SF probability $P_S (E)$ to that obtained from the Elliott relation. At some energies, e.g., 0.5\,-\,1 eV, these two quantities are nearly the same, but at other energies there is no direct relation other than that SF probability is large where band states are present.
An interesting difference in the context of ultrafast demagnetization is the suppression
of  $P_{S} (E)$ around $E_F$, which is not captured by $P_{S}^{b^{2}}(E)$. 
 The features of $P_{S}(E)$ that are not captured by $P_{S}^{b^{2}}(E)$
 can be understood by comparing Eqs.\ (\ref{eq:SFElbgEn}) and (\ref{eq:EllRelDOS}).
 One of the differences is the presence/absence of summation over destination
eigenstates $\mathbf{k}'n'$.  The latter are restricted in Eq.\ (\ref{eq:SFElbgEn})
by the construction of $g_{\mathbf{k}n,\mathbf{k}'n'}^{\nu\uparrow \downarrow}$ to correspond
to a {\it different} spin than the source state $\mathbf{k}n$. The number
of available end states is however not taken into account in Elliott formula
(which, derived for a paramagnetic metal, assumes that the same number of states is available for both spins, and hence suppresses this distinction).
The mentioned discrepancy between $P_{S}$
and $P_{S}^{b^{2}}$ above $E_F$ is thus easily explained by
the lack of states with the same energy and opposite spin in the Ni
DOS (see Fig.\ \ref{fig:TransScheme}). Hence, the Elliott relation fails for ferromagnets in strongly exchange-split energy regions. 
 
%

After laser-excitation electrons equilibrate quickly due to electron-electron scattering at a high electron temperature $T_e$ of the order of thousands K. 
To describe this situation we use appropriate $f_{\sigma}(E)$, but note that the chemical potential must be adjusted also. 
Spin conservation leads to differences between $f_{\uparrow} (E)$ and $f_{\downarrow} (E)$, namely $f_{\downarrow} (E)$ has a lower chemical potential than $f_{\uparrow} (E)$ in Ni due to the shape of its DOS.
SF probabilities $P_S$ computed for several $T_e$ are given in Table \ref{tab:SFRates-1}. With increasing $T_e$ $P_S$ increases, too. Also the Elliott SF probability $P_{S}^{b^{2}}$ increases with $T_e$, but it deviates still from $P_S$. 
A previous work \cite{koopmans10} used a Gaussian smearing to stimulate a thermalized system (without $E_F$ adjustment) and obtained $P_{S}^{b^{2}} \approx$0.18. Our values are smaller, but note that the way the thermalized distribution is described is different.

As mentioned before, a large SF probability does not necessarily imply a large demagnetization.
Evaluating the demagnetization rate $dM/dt$=$2\mu_B (S^-\!-\!S^+)$ for thermalized electron distributions we obtain quite small values, of the order of 0.08$\mu_B$/ps. 
The reason is that not just a large SF probability, but also an imbalance between $f_{\uparrow}(E)$ and $f_{\downarrow}(E)$ is essential for a magnetization change. 
The distributions of spin populations specific to Ni 
imply that for thermalized electrons
below $E_F$ most spin-flips {\it increase} the spin moment,
spin-reducing transitions occur only above $E_F$. 
In that region the SF scattering rate is however very low (Fig.\ \ref{fig:En_SF}). 
The situation is illustrated in Fig.\ \ref{fig:TransScheme}. 
As a consequence the spin-decreasing rate ($S^-\!-\!S^+$) is thus much lower than the SF rate ($S^-\!+\!S^+$),
and in addition it exhibits only a weak temperature
dependence. 
Hence we find that phonon-mediated SF scattering in thermalized Ni cannot be the mechanism of the observed ultrafast demagnetization.

%

One remaining possibility for a fast demagnetization is an enhanced SF rate in the NEQ distribution present immediately after the laser pulse. 
Previous {\it ab initio} calculations showed that minority-spin electrons are excited more than majority-spin ones, see \cite{oppeneer04}.
Assuming a 1.5-eV pump-laser and a simplified step-like electron distribution 
reduced by about 5\% 
in the 1.5-eV energy window below $E_F$, the calculated demagnetization ratio $D_S$ is higher than for thermalized distributions (Table \ref{tab:SFRates-1}).
A critical role is played here by holes deep below $E_F$ with high SF probability as well as a significant difference between majority and minority occupations (see Fig.\ \ref{fig:TransScheme}).
An important yet unknown element in estimating the demagnetization is the laser fluence.
Nonetheless, we find that phonon-mediated demagnetization in Ni is much more effective in the NEQ state than in the thermalized state, as was proposed recently for Gd \cite{wietstruk11}.
An important aspect is the time scale on which the NEQ demagnetization is active.
Electron thermalization proceeds fast in Ni and transforms the initial NEQ distribution to a thermalized one in $\sim$200 fs. A rough estimate of the demagnetization in this time-window is 0.1$\mu_B$, i.e. smaller than the observed experimental demagnetization. The precise amount of the demagnetization depends however on the time-evolution of the distributions, which requires further investigations. 

Using relativistic {\it ab initio} calculations we have evaluated the phonon-induced SF probability
and demagnetization in laser-pumped Ni. A strong dependence of these quantities on the electron energy is observed, which is not tracked by the Elliott approximation. In the electron thermalized state Elliott-Yafet phonon-mediated demagnetization is too small to explain the ultrafast demagnetization, despite reasonably large SF probabilities. We find that Elliott-Yafet SF scattering contributes more to the demagnetization for NEQ distributions immediately after the fs laser-excitation.
We note lastly that the existence of {\it other} fast SF channels \cite{carpene08,krauss09,bigot09,schmidt10} cannot be excluded.

\acknowledgments
{We thank H.C.\ Schneider, J.\,K.\ Dewhurst and Th.\ Rasing for valuable discussions. 
This work has been supported by the Swedish Research Council (VR), by FP7
EU-ITN ``FANTOMAS",  the G.\ Gustafsson Foundation, Czech Science Foundation (P204/11/P481), 
and the Swedish National  Infrastructure for Computing (SNIC).}

\end{document}